\documentclass{kapproc} % Computer Modern font calls

\usepackage{procps} 

\usepackage[dvips]{graphicx}

\upperandlowercase

\kluwerbib % will produce this kind of bibliography entry:

\newcommand{\HII}{\mbox{H\,{\sc ii}}}
\newcommand{\HI}{\mbox{H\,{\sc i}}}

\newcommand{\Htwo}{H$_{2}$}

\newcommand{\pcmcub}{\mbox{${\rm cm^{-3}}$}}

\newcommand{\pcmsq}{\mbox{${\rm cm^{-2}}$}}
\newcommand{\pwr}[2]{\mbox{$#1 \times 10^{#2}$}}

\newcommand{\Gzero}{$G_0$}

\newcommand{\lsim}{\mbox{$\mathrel{\vcenter{\hbox{\ooalign{\raise3pt\hbox{$<$}\crcr \lower3pt\hbox{$\sim$}}}}}$}}
\newcommand{\gsim}{\mbox{$\mathrel{\vcenter{\hbox{\ooalign{\raise3pt\hbox{$>$}\crcr \lower3pt\hbox{$\sim$}}}}}$}}

\usepackage{amssymb}	%needed for "\lesssim" & "\gtrsim"
\begin{document}

\articletitle[Photodissociation and the Morphology of \HI\ in Galaxies]
{Photodissociation and the \\
Morphology of \HI\ in Galaxies}

\author{Ronald J. Allen} 
 
\affil{Space Telescope Science Institute \\
3700 San Martin Drive, Baltimore MD 21218, USA}

\begin{abstract}
Young massive stars produce Far-UV photons which dissociate the molecular gas
on the surfaces of their parent molecular clouds. Of the many dissociation
products which result from this ``back-reaction'', atomic hydrogen \HI\ is one
of the easiest to observe through its radio 21-cm hyperfine line emission. In
this paper I first review the physics of this process and describe a
simplified model which has been developed to permit an approximate computation
of the column density of photodissociated \HI\ which appears on the surfaces
of molecular clouds. I then review several features of the \HI\ morphology of galaxies on a variety of length
scales and describe how photodissociation might account for
some of these observations.
Finally, I discuss several consequences which follow if this view of the
origin of HI in galaxies continues to be successful.
\end{abstract}


\section{Introduction}

I want to begin this talk with a brief discussion of some aspects of the
astrophysics of photodissociation regions (PDRs) in
order to make it clear that the production of \HI\ from \Htwo\ in the
ISM is quite inevitable, and that the cycling of \HI\ $\leftrightarrow$ \Htwo\
is likely to be both continuous and ubiquitous in galaxies. I will then move on
to a brief description of some of the features of the \HI\ morphology of galaxies
which appear to be amenable to an explanation in terms of PDR astrophysics.

The material presented here is a review of results and views published
by myself and by others in previous journal and conference papers. New in this
paper are some preliminary results on the effects of radial
gradients in the metallicity of the ISM on the overall \HI\ distribution
in galaxies in the context of the photodissociation picture described here. 
 
\section{\HI\ from photodissociation of \Htwo\ in the ISM}
\label{sec:HI}

The physics and astronomy of the photodissociation $\leftrightarrow$ reformation
process for molecules in the ISM is a subject of active research, and an excellent
review of the field was published a few years ago by \cite{hol99}. The results
have been successfully applied to star-forming regions in the Galaxy on length
scales of typically 0.1 - 1 pc.
The first indication that photodissociation may be operating to affect
the \textit{large-scale} morphology (100-1000 pc) of the \HI\ in galaxies was
found in M83 by \cite{all86}, who noticed a clear
spatial separation between a particularly well-defined dust lane several
kiloparsec long and
the associated ridge of \HI\ and \HII\ in the southern spiral M\,83.
Appropriately for this conference, M\,83 is a barred spiral galaxy, and the
strong stellar density wave which is apparently driven by the bar has produced
large streaming velocities in the gas across the arms, leading to a measureable
separation of various phases of the ISM.

\subsection{Dissociation of \Htwo\ by far-UV photons}

It requires a photon of energy $\geq 14.67$ eV to lift an \Htwo\ molecule
directly from its ground state to the electronic continuum. Such photons will
be rare in the ISM since they are strongly absorbed in ionizing \HI\ to \HII.
For this reason it was initially thought (\cite{spi48}) that \Htwo\ would be very
long-lived in the ISM. However, closer examination of the electronic and
vibrational energy-level diagram for \Htwo\ revealed another dissociation
channel. In this ``fluorescence'' process, less energetic photons
can raise the \Htwo\
molecule to an excited electronic state. When the molecule decays to the ground
electronic state, it can end up in a variety of excited vibrational states.
However, vibrational levels above 14 are so energetic that they break the
chemical bond holding the \Htwo\ molecule together, and two \HI\ atoms are
produced.
A quantitative description of this process was first given by \cite{ste67}.
It starts when \Htwo\ molecules absorb photons primarily at wavelengths of
$\lambda \approx 110.8$ and $\approx 100.8$ nm
through transitions to the to the electronically-excited Lyman
($X ^1\Sigma_g^+ \rightarrow B^1\Sigma_u^+$) and Werner
($X ^1\Sigma_g^+ \rightarrow C ^1\Pi_u^+ $) bands. In the subsequent decay
to various vibrationally-excited levels of the ground electronic
state, $\sim 10 - 15$\%
of the \Htwo\ molecules will dissociate into two \HI\ atoms. Considering that
even higher electronic states exist in \Htwo, it is clear that the FUV spectrum
over the whole range from 91.2--110.8 nm (13.6--11.2 eV) contributes to
the dissociation. In high-UV-flux
environments (\Gzero\ $\gtrsim 10^4$, \cite{shu78}), photons with wavelengths
as long as $\approx 185$ nm ($\approx 6.6$ eV) can continue to create \HI\
by dissociating ``pumped'' \Htwo\ ($X^1\Sigma_g^+, 2 < v \leq 14$) via additional
Lyman- and Werner-band transitions.
Verification that this process actually occurs in the ISM was
obtained when the predicted UV fluorescence spectrum was first observed by
\cite{wit89} in the Galactic nebula IC 63.

Using the notation of \cite{ste88}, the rate at which \Htwo\
is dissociated by this process in the ISM per unit volume can be written as:
\begin{equation}
R_{diss} = D \chi \times e^{-\tau_{gr,1000}} \times f_s(N_2) \times n_2,
\end{equation}
in units of \Htwo\ molecules dissociated \pcmcub\ sec$^{-1}$,
where $n_{2} = n$(\Htwo), and \\
\\
\begin{tabular}{rcl}
$D$ & = & the unattenuated \Htwo\ photodissociation rate in the average ISRF, \\
$\chi$ & = & the incident UV intensity scaling factor, \\
$\sigma$ & = & the effective grain absorption cross section per H nucleus \\
 & & in the FUV continuum, \\
$\tau_{gr,1000}$ & = & $\sigma (N_{1} + 2N_{2})$ is the dust grain opacity at
$\lambda \sim 100$ nm, and \\  
$n$ & = & $n_{1} + 2n_{2}$ the volume density of H nuclei. \\
\end{tabular}
\\

\noindent Since $D = 5.43 \times 10^{-11}$ s$^{-1}$ (according to the
simplified 3-level model, \cite{ste88}) the time scale for this process on the
surface of a typical GMC ($n_{2} \sim 50$ molecules \pcmcub) illuminated by
the ISRF is $\tau_{diss} = (D n_{2})^{-1} \approx 10$ yr! Note that each dissociation produces two \HI\ atoms on the cloud surface, and the appearance
of a layer of \HI\ when a FUV radiation field is ``switched on'' (e.g.\ from
the birth of a new O--B star) is instantaneous compared to most other
time scales in the ISM. 

\subsection{Formation of \Htwo\ on dust grains}

Formation of \Htwo\ occurs in the ISM most efficiently on dust grains
(cf.\ e.g.\ \cite{hol99} and references cited there).
The model for the formation rate depends on
several parameters (some of which are not accurately known), as well as on the
nature of the dust grains (which may vary from place to place in a galaxy). The
usual parametrization is:
\begin{equation}
R_{form} = \gamma_{2} \times n \times n_{1}
\end{equation}
in units of \Htwo\ molecules \pcmcub\ sec$^{-1}$, and $n = n_{1} + 2n_{2}$.
The rate coefficient for
unit density, solar metallicity, and 100K kinetic temperature (roughly the ISM
in the solar neighborhood)
is $\sim$ \pwr{1-3}{-17}\ cm$^{3}$ sec$^{-1}$. This equation
is strongly dependent upon the dust--to--gas ratio
($\delta=A_V/N_H$) and weakly dependent upon the gas temperature ($T$), since
\begin{equation}
\gamma_{2} = 3.0 \times 10^{-18} (\delta/\delta_0) T^{1/2}
y_F(T)\ {\rm cm^3\ s^{-1}}, \nonumber
\end{equation}
where $\delta_0$ refers to the value of $\delta$ in the solar neighborhood,
and $y_F(T)$ represents the efficiency of \Htwo\ formation.  The
product $T^{1/2} y_F(T)$ is thought to be constant to within a factor of 2
(\cite{hol71}).

The time scale for this process is $\tau_{form} = (2n\gamma_{2})^{-1} \approx$
\pwr{5}{8}/$n$ yr, and this will be the rate-determining time scale in an
equilibrium situation where photodissociation is
balanced by reformation on dust grains.

\subsection{Equilibrium}

In recent years, much effort has
gone into calculating the level populations of the ro-vibrational lines of the
ground state of \Htwo, and the intensities of the associated
quadrupole line emission. These lines can be observed with space missions (e.g.
SWAS, ISO) in PDRs in the Galaxy and in the nuclear regions of other galaxies.
The \HI\ column density in a PDR is calculated with the same physics used to
determine the excitation of the \Htwo\ near-infrared fluorescence lines. While
the computations for those lines are rather complicated, the determination of
\HI\ column can be obtained from a simplified version of the model. Furthermore
the 21-cm line emission from \HI\ is almost always optically thin, so the
observations usually yield the \HI\ column directly for comparison with the model. 
The
formation and destruction rates described above are set equal, and the equation
solved, in the present case for the \HI\ column density. A logarithmic form for
the analytic solution to this equation was first given by \cite{ste88}; see also Appendix A of this paper for a brief derivation. Other
relevant references are given in \cite{all04}.
The model is a simple semi-infinite slab geometry in
statistical equilibrium with FUV radiation incident on one side.
The solution
gives the steady state \HI\ column density along a line of sight perpendicular
to the face of the slab as a function of $\chi$, the incident UV intensity
scaling factor, and the
total volume density $n$ of H nuclei.  Sternberg's result is:
\begin{equation}
N(\HI)={1 \over \sigma} \times \ln{\left[ {{D \mathcal{G}} \over
{\gamma_{2} n}}\chi + 1 \right]},
\label{eqn:nhi}
\end{equation}
%\noindent where;

\begin{tabular}{rcl}
$D$ & = & the unattenuated \Htwo\ photodissociation rate in the average ISRF, \\
$\gamma_{2}$ & = & the \Htwo\ formation rate coefficient on grain surfaces, \\
$\sigma$ & = & the effective grain absorption cross section \\
   &   & per H nucleus in the FUV continuum, \\
$\chi$ & = & the incident UV intensity scaling factor, \\
$N(\HI)$ & = & the \HI\ column density, \\
$n$ & = & the volume density of H nuclei. \\
\end{tabular}
\\

Equation \ref{eqn:nhi} has been developed using a simplified three-level model
for the excitation of the \Htwo\ molecule and is applicable for low-density
($n \lesssim 10^4$ \pcmcub), cold (T $\lesssim 500$ K), isothermal, and static
conditions, and neglects contributions to $N(\HI)$ from ion chemistry and
direct dissociation by cosmic rays.  The quantity $\mathcal{G}$ here
(not to be confused
with \Gzero\ to be defined momentarily) is a dimensionless function of the effective
grain absorption cross section $\sigma$, the absorption self--shielding
function $f_{s}(N_{2})$, and the column density of molecular hydrogen $N_2$:
\begin{displaymath}
\mathcal{G} = \int_0^{N_2} \sigma f_{s} e^{-2\sigma N_2^\prime} {\rm d}N_2^\prime.
\end{displaymath}
The function $\mathcal{G}$ becomes a constant for large values of $N_2$ due to
self--shielding (\cite{ste88}).  Using the parameter values in this equation
adopted by \cite{mad93}, we have:
\begin{displaymath}
N(\HI) = \pwr{5}{20} \times \ln [1+ (90\chi/n)]
\end{displaymath}
\noindent where $n$ is in \pcmcub.  This is a steady state model, with \Htwo\
continually forming from \HI\ on dust grain surfaces, and \HI\ continually
forming from \Htwo\ by photodissociation.

Equation \ref{eqn:nhi} is strongly dependent upon the dust--to--gas ratio
($\delta=A_V/N_H$) and weakly dependent upon the gas temperature ($T$), since
\begin{eqnarray}
\sigma & = & 1.883 \times 10^{-21} (\delta/\delta_0)\ {\rm cm}^{-2}, \nonumber\\
\gamma_{2}  & = & 3.0 \times 10^{-18} (\delta/\delta_0) T^{1/2} y_F(T)\
{\rm cm^3\ s^{-1}}, \nonumber\\
\mathcal{G} & = & (\sigma/\sigma_0)^{1/2}\mathcal{G_{O}}, \nonumber
\end{eqnarray}
where $\delta_0$, $\sigma_0$, and $\mathcal{G_{O}}$ refer to values in the solar neighborhood,
and $y_F(T)$ represents the efficiency of \Htwo\ formation.  The
product $T^{1/2} y_F(T)$ is thought to be constant to within a factor of 2
(\cite{hol71}).

Equation \ref{eqn:nhi} also contains a dependence on the level of
obscuration in the immediate vicinity of the FUV source.  While the
variable $\chi$ represents the intrinsic FUV flux associated with the
star-forming region, we generally observe an attenuated FUV
flux. Assuming any extinction associated with the star-forming region
is in the form of an overlying screen of optical depth $\tau (FUV)$,
$\chi({\rm observed})=\chi e^{-\tau (FUV)}$. However, an accurate
correction for this effect may be difficult, Since the ISM in the
immediate vicinity of the FUV source will have been disturbed by
stellar winds and any prior supernovae.

Assuming solar neighborhood values of $\sigma_0=1.883 \times 10^{-21}$
cm$^2$, $D=5.43 \times 10^{-11}$ s$^{-1}$, $\gamma_{20}=3 \times 10^{-17}$
cm$^{3}$ s$^{-1}$, and $\mathcal{G_{O}} \approx 5 \times 10^{-5}$
(\cite{ste88}), and neglecting the weak
temperature dependence of $\gamma_{2}$, equation \ref{eqn:nhi} becomes:
\begin{equation}
N(\HI)= {{5 \times 10^{20}} \over {(\delta/\delta_0)}}
\ln{\left[ {{90 \chi } \over {n}}
{{\left( {{\delta} \over {\delta_0}} \right)}^{-1/2}}
+ 1 \right]}, 
\label{eqn:nhidel}
\end{equation}
where $\chi=\chi({\rm observed})e^{\tau (FUV)}$.
% ++++++++++++++++++++++++++++++++++++++++++++++++++++++++++++++++++++++
\begin{figure}[ht]
\vskip0.0in
\centerline{\includegraphics[width=4in]{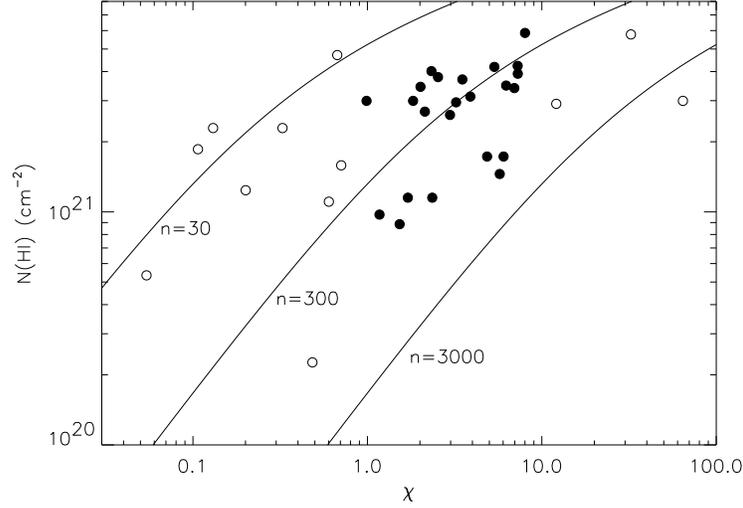}}
\caption{The Relationship between $N(\HI)$ and
$\chi$.  The observed and predicted values of $N(\HI)$ are shown as a
function of $\chi$. The modeled behavior of $N(\HI)$ assumes values of
$\delta/\delta_0=0.2$ and $\tau (FUV)=0$, appropriate for the outer regions
of M101.  The observations are clearly consistent with the physics
underlying the photodissociation picture. \label{fig:nhichi}}
\end{figure}
% ++++++++++++++++++++++++++++++++++++++++++++++++++++++++++++++++++++++
The behavior of $N(\HI)$ as a function of $\chi$ is displayed for
$\delta/\delta_0=0.2$, $\tau (FUV) = 0$ and $n=30$, $n=300$, and
$n=3000$ in Figure \ref{fig:nhichi}.  These values of $\delta/\delta_0$ 
and $\tau (FUV)$ are appropriate for the outer regions of M101, as 
discussed in \cite{smi00}. Values of $N(\HI) \gtrsim
10^{22}$ cm$^{-2}$ are not likely to be observed as the atomic gas
probably becomes optically thick at this point:
\begin{eqnarray}
N(\HI)	& = & 1.82 \times 10^{18} \int_{- \infty}^\infty T_s \tau (v)
dv \nonumber \\
 & \rightarrow & 1.82 \times 10^{18} T_s \tau \Delta v \nonumber \\
 & \approx & 10^{22} {\rm cm}^{-2}, \nonumber
\end{eqnarray}
for spin temperatures of $T_s \approx 100$K and profile FWHMs of
$\Delta v \approx 20$ km s$^{-1}$ typical of M101 (\cite{bra97}), and for
optical depths of $\tau \approx 2.5$, corresponding to a ratio between
the brightness and kinetic temperatures of $T_B/T_K \approx 0.9$.
This value is appropriate for the highest-brightness regions of M101, as
indicated in Figure 8a of \cite{bra97}.  Figure \ref{fig:nhichi} also
shows the measurements for each of the 35 candidate PDRs in M101 as analysed
by \cite{smi00}.  The data
indicate that the properties of observed regions in M101 are
consistent with photodissociation of an underlying molecular gas of
moderate volume density.

\cite{all04} have recently re-examined equation \ref{eqn:nhidel} and compared it
with the full numerical treatment used in the ``standard'' Ames model
summarized in \cite{kau99}. A conversion of the FUV flux
$\chi$ used by \cite{ste88} to the quantity
\Gzero\ used by \cite{kau99} is first required; this is because
\cite{ste88} and \cite{kau99}
use different normalisations for the FUV flux (see Appendix B in \cite{all04}).
When distributed sources illuminate an FUV-opaque PDR over $2\pi$ sr, the
conversion is $\chi = G_0/0.85$ (see Footnote 7 in \cite{hol99}), resulting
in $90 \chi /n = 106 G_0/n$. With this change, \cite{all04} fitted the
analytic expression for $N(\HI)$ above to the model computations in
the range in which cosmic ray dissociation is not a major contributor, roughly for
\Gzero\ $\gtrsim 1$, $n \gtrsim 10$ \pcmcub.  The result is that no consistent
improvement is obtained by using any value for the coefficient of \Gzero\ other
than the value 106 deduced above, although a modest improvement is obtained by
using a slightly larger value for the leading coefficient in the equation,
\pwr{7.8}{20}, corresponding to a value of \pwr{1.3}{-21} cm$^2$ for the
effective grain absorption cross section.  With these small adjustments, the
final best-fit equation is:
\begin{equation}
N(\HI)= {{7.8 \times 10^{20}} \over {(\delta/\delta_0)}}
\ln{\left[ {{106G_0} \over {n}}
{{\left( {{\delta} \over {\delta_0}} \right)}^{-1/2}}
+ 1 \right]} {\rm ~cm}^{-2}, 
\label{eqn:dissociate}
\end{equation}
where $n = n(HI) + 2n(H_{2})$ and \Gzero\ and $\delta_{0}$ are the (normalized)
FUV flux and dust/gas ratio in the ISM of the solar neighborhood.
% ++++++++++++++++++++++++++++++++++++++++++++++++++++++++++++++++++++++
\begin{figure}[ht]
\vskip0.0in
\centerline{\includegraphics[width=4in,height=3in]{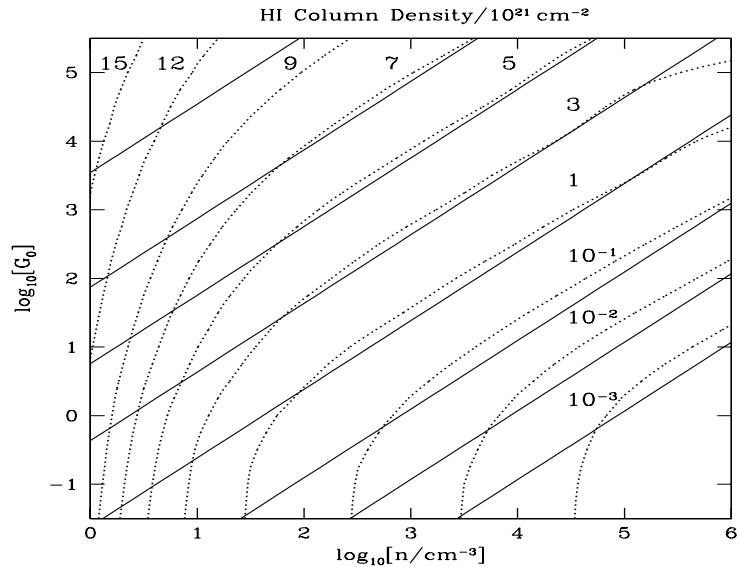}}
\caption {Contours of constant \HI\ column density $N(\HI)$ in units of
$10^{21}$ \pcmsq\ in the PDR as a function of the density $n$ and incident FUV
flux \Gzero\ for the standard model parameters in \cite{kau99}
(\textit{dotted lines}), and for the analytic approximation of eq.\
\ref{eqn:dissociate} (\textit{solid lines}).  The labeled contour values are
for the numerical model; the contours for the analytic model start from $10^{-3}$
(lower right corner) and increase to $10^{-2}$, $10^{-1}$, 1, 3, 5, 7, and end
at 10 (upper left corner).  \label{fig:modelHIboth}}
\end{figure}
% ++++++++++++++++++++++++++++++++++++++++++++++++++++++++++++++++++++++
In Figure \ref{fig:modelHIboth} we show values from equation
\ref{eqn:dissociate} plotted as solid lines together with dotted contour
lines from the ``standard'' numerical model for $\delta = \delta_{0}$.
The agreement is generally good over
much of the $n$--\Gzero\ parameter space of interest here; differences occur
mainly in the top left corner of the diagram, and at low values of FUV flux.
In the top left corner the analytic formula under-predicts the amount of \HI\
column density computed from the standard model by about 30\% owing to \HI\
production by ion chemistry reactions such as $H_2^{+} + H_2 \rightarrow
H_3^{+} + H$, $HCO^{+} + e^{-} \rightarrow CO + H$, and $PAH^{-} + H^{+}
\rightarrow PAH + H$ (where PAH is a polycyclic aromatic hydrocarbon), which
are important at high \Gzero\ and low $n$.  At values of \Gzero\ $\lesssim 1$
the contours of $N(\HI)$ for the numerical model become vertical; this is
because the standard model includes a low level of cosmic ray ionization
which contributes a small amount of \HI\ by dissociation of \Htwo\ even
for \Gzero\ = 0.

It should be noted that
both the analytic and the numerical forms depend on several rather crude
parameters used to describe the properties of the gas -- dust mixture in the
ISM (dust cross section, \HI\ sticking coefficients, etc.), and that not all
proponents of PDR models use the same values for these parameters. A coordinated
effort is presently taking place among the modellers, first to see if different
PDR codes can produce the same results when using the same numerical values for
parameters (no, they can differ, and sometimes by a lot!), and second,
to try to reach some
agreement on an acceptable set of values for these parameters.

There are several noteworthy aspects of this equation:
\begin{itemize}
\item $N(\HI)$ depends only on the \textit{ratio} of $G_0 / n$. Low FUV
flux, low density environments in galaxies can produce the same column of \HI\
found in high flux, high density environments. The difference will be in the
thickness of the \HI\ layer; much thicker layers of \HI\ are associated with
the low flux, low density environments;
\item at a given $n$, $N(\HI)$ increases first linearly with $G_0$, but then ``saturates'' and increases only \textit{logarithmically} after that;
\item at a given $G_0$, $N(\HI)$ \textit{decreases} logarithmically with increasing $n$, and;
\item the \HI\ column \textit{decreases} as the dust/gas ratio increases.
\end{itemize}

\section{Time scales}

As discussed in the previous section, the relevant time scale for the production
of \HI\ on the surfaces of molecular clouds is the formation time for \Htwo\
on dust grains. The full expression including the dust/gas dependence is:
\begin{eqnarray}
\tau_{2} & = & (2n\gamma_{2})^{-1} \nonumber \\
    & = & \frac{5 \times 10^{8}}{(n_{1} + 2n_{2})\times(\delta/\delta_{0})} 
    \ \mbox{\rm yrs.}
\end{eqnarray}
Table \ref{table:times} gives typical values for $\tau_{2}$ in several different
environments in the ISM, and Table \ref{table:othertimes} lists a number of time
scales set by other processes in galaxy disks.

\begin{table}[ht]
\caption[]%<-- this version will appear in List of Tables
{\HI\ production time scales in typical ISM environments. \label{table:times}}
%<-- this version will appear on page
\begin{tabular*}{\textwidth}{@{\extracolsep{\fill}}lccc}
\sphline
Environment & $n$ & $\delta/\delta_{0}$ & $\tau$ (years) \\
\sphline
Solar metallicity GMC & 100 & 1 & $\sim 3 \times 10^{6}$ \\
Intercloud gas & 10 & 1 & $\sim 3 \times 10^{7}$ \\
Outer galaxy GMC & 100 & 0.1 & $\sim 3 \times 10^{7}$ \\
\sphline
\end{tabular*}
\end{table}
%\inxx{captions,table}
\begin{table}[ht]
\caption[]%<-- this version will appear in List of Tables
{Other time scales in galaxies. \label{table:othertimes}}
%<-- this version will appear on page
\begin{tabular*}{\textwidth}{@{\extracolsep{\fill}}lc}
\sphline
Situation & Time scale (years) \\
\sphline
GMC crossing time $R/\Delta V$ & $\sim 7 \times 10^{6}$ \\
Spiral arm crossing time & $\sim 5 \times 10^{7}$ \\
B3 star lifetime for FUV production & $\sim 5 \times 10^{7}$ \\
Galaxy rotation time & $\sim 5 \times 10^{8}$ \\
Hubble time & $\sim 10^{10}$ \\
\sphline
\end{tabular*}
\end{table}
%\inxx{captions,table}
It is clear that the time scale for \HI\ production by photodissociation of \Htwo\
is short enough to be relevant. For instance, \HI\ is produced in the same time
scales as that of molecular cloud formation (\cite{pri01}) and of the star
formation in those molecular clouds (\cite{elm00}), and in a tenth of the lifetime
of a typical B star for FUV photon producation. This must be the reason why we
see \HI\ in regions where young stars form, and why \HI\ appears in ``rims''
around clusters of B stars in spiral arms. \HI\ also appears in a time short
compared to the time it takes for a GMC to cross a spiral arm, and the \HI\ will
correspondingly disappear as the FUV production ceases further down stream and the
gas reverts to a predominantly molecular form. Even in the far outer parts of galaxies we see that the \HI\ time scale is still only about 10\%
of the rotation time of the galaxy, so for undisturbed galaxies we can expect
an equilibrium to obtain between \HI\ and \Htwo\ even in these sparse environments.

\section{Major features in the \HI\ morphology of galaxies}
\label{sec:obs}

Our picture of the major features of the \HI\ morphology in disk galaxies
has closely followed the steady improvements in angular resolution of
centimeter-wave radio telescopes, first with filled apertures
(Dwingeloo, NRAO 300', Parkes, Arecibo, GBT...), and later with synthesis
imaging instruments (Westerbork, VLA, ATNF...).

\subsection{Radial distribution of \HI\ surface brightness}
% ++++++++++++++++++++++++++++++++++++++++++++++++++++++++++++++++++++++
\begin{figure}[ht]
\vskip0.0in
\centerline{\includegraphics[width=4in,height=3in]{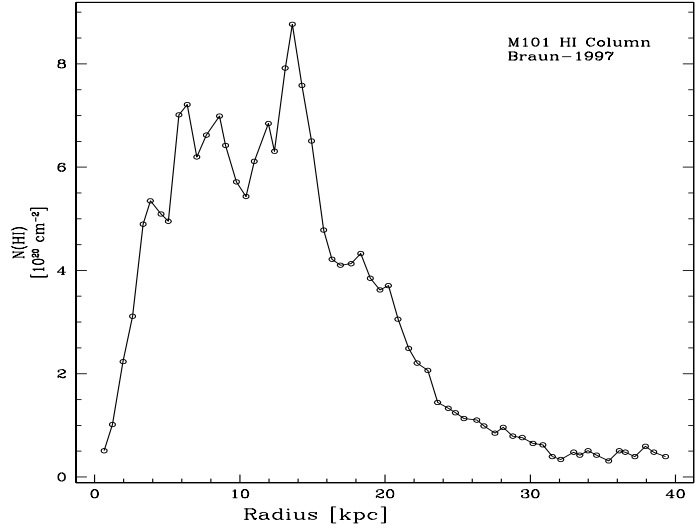}}
\caption{Radial distribution of \HI\ surface brightness for
the nearby giant Sc galaxy NGC5457 = M\,101 obtained by averaging the
\HI\ data in annular elliptical rings. From \cite{bra97}, adjusted to an assumed
distance of 5.4 Mpc. $R_{25}$ for this galaxy is $13.5' \approx 21$ kpc. \label{fig:radialHI}}
\end{figure}
% ++++++++++++++++++++++++++++++++++++++++++++++++++++++++++++++++++++++
A typical averaged \HI\ radial surface brightness profile of a nearby
giant Sc galaxy is shown in Figure \ref{fig:radialHI}. The main features
to notice are:
\begin{itemize}
\item the central depression; 
\item the ``flat top'', typically at a level of \pwr{5 - 10}{20} \pcmsq, and;
\item the long ``tail'' to faint levels in the distant outer parts.
\end{itemize}

What parts of this plot can be explained by the photodissociation picture
descibed in the previous section? An explanation for the ``flat top'' was
actually suggested in terms of the photodissociation picture nearly 20 years
ago by \cite{sha87}, but I fear that paper has been widely ignored by most
workers in the field of extragalactic \HI\ and \Htwo. A good look at equation
\ref{eqn:dissociate} makes it clear: owing to the \textit{logarithmic}
dependence of N(\HI) on the \textit{ratio} \Gzero /$n$ and the fact that
the largest concentrations of young stars also go along with regions of highest
gas density, we ought not to expect values of \HI\ column density
much in excess of a few times the coefficient in front of the log.
So even in the most active starbursting regions of galaxies, we are not
likely to observe values of the \HI\ column density much in excess of
$\sim$ \pwr{few}{21} \pcmsq. As to the decline in the outer regions, this
is likely to be a combined effect of a declining ambient FUV flux and a declining
area filling factor, since \cite{smi00} have shown that the total gas density $n$
on the surfaces of GMCs located in the neighborhood of massive
young stars does not appear to change much with radius.

\begin{figure}[ht]
\vskip0.0in
\centerline{\includegraphics[width=4in,height=3in]{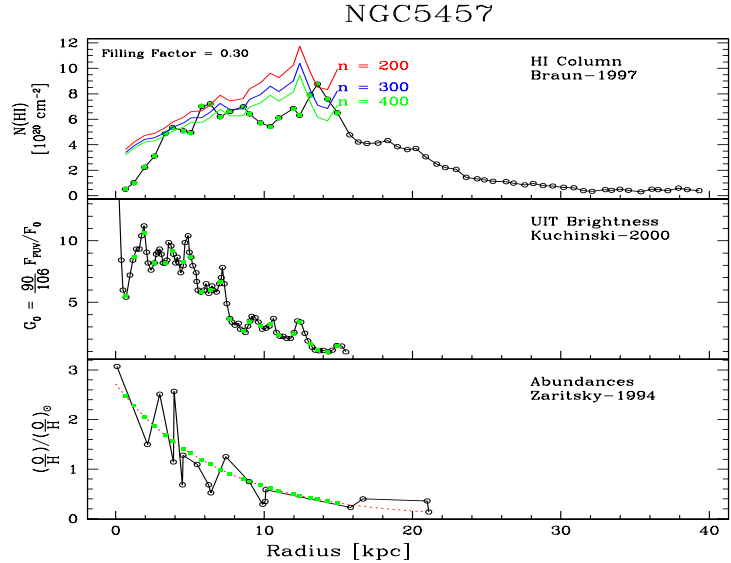}}
\caption {\HI\ column density (top), FUV brightness (middle), and O/H ratio (lower) as a function of radius for M\,101. Equation \ref{eqn:dissociate} has been used
to produce the fits shown in the top panel for an area filling factor of 0.3 and
$n = 100, 200, \&\ 300$ \pcmcub. From work in progress by Allen, Waghorn, \& Heiner.
\label{fig:M101-HI-radius-plot}}
\end{figure}

Finally, the depression in the inner parts can be
explained as a consequence of the general increase in the metallicity of the
ISM in the inner parts of galaxy disks. In Figure \ref{fig:M101-HI-radius-plot}
my student colleague Ben Waghorn has fitted equation \ref{eqn:dissociate} to the
combined radial data on the FUV distribution and the metallicity gradient in M\,101.
The free parameters are the \HI\ area filling factor (taken here to be 0.3
everywhere), and the GMC gas volume density $n$ (fits shown for 100, 200, and
300 \pcmcub). We see that, in spite of the
strong increase in FUV flux \Gzero\ in the inner parts of the galaxy, the rapid
rise of the dust/gas ratio (assumed proportional to the O/H ratio) actually
results in a \textit{decrease} in N(\HI), as observed, and the quantitative fit
is also reasonable. I note here that this result has already been described
by \cite{smi00} for a small subset of young star clusters in M\,101 accounting
for only a few percent of the total \HI\ content of the galaxy; what we are now
seeing is that the same explanation appears viable for \textit{all} the \HI\
in the galaxy. We have examined nearly a dozen nearby spirals in this way, and
find reasonable agreement for about half of them. The other half show an
indication that there is more \HI\ present than FUV-related photodissociation
can explain. Interestingly, these galaxies nearly all have bright nonthermal radio
continuum disks, suggesting that there is a component of the \HI\ being
maintained from dissociation by cosmic rays penetrating throughout the GMCs.
This work is ongoing.

% ++++++++++++++++++++++++++++++++++++++++++++++++++++++++++++++++++++++
% ++++++++++++++++++++++++++++++++++++++++++++++++++++++++++++++++++++++

\subsection{Spiral structure}

In the Introduction to this paper I pointed out that it was thanks to a
strong density wave in the southern barred spiral M\,83 that the importance
of photodissociation in affecting the morphology of galaxies on the large
scale was first unmasked. Other studies have
followed on M83 and on other galaxies (M\,51, M\,100; see \cite{smi00} for references) and have generally agreed that the initial
interpretation in terms of photodissociation remains the most viable option.
The separation in the case of M\,83 is about 250 pc, and arises because
of the difference between the spiral pattern speed and the rotation
speed of the gas, coupled with the time for collapse of GMCs and the
time that a massive young star lives on the main sequence.

\begin{figure}[ht]
\vskip0.0in
\centerline{\includegraphics[width=4in]{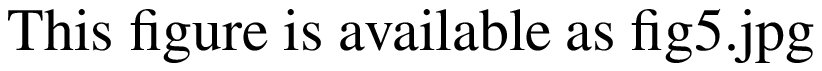}}
\caption{The distribution of atomic hydrogen as seen at a resolution
of $6'' \approx 200$ pc in M\,101 (\cite{bra95}). Thin, prominent spiral
arms wind outwards to beyond the optical disk at $R_{25} \approx 13.5'$,
with the outer \HI\ arms generally traceable back to
arms which start in the main body of the disk. Contours of the Far-UV
emission recorded by UIT are superposed. From the PDR analysis by \cite{smi00}.
\label{fig:M101-HI}}
\end{figure}

The clear existence of spiral features in the \HI\ distribution of a galaxy
external to our own was perhaps first convincingly demonstrated for M\,101
by \cite{all73}. Although this galaxy apparently does not have a very strong
density wave, the \HI\ is arranged in thin spiral segments which appear in the
inner disk and can be traced over a large part of the main body of the galaxy right out to beyond $R_{25}$.
Figure \ref{fig:M101-HI} shows the \HI\ image (grey, kindly provided
in digital form by R. Braun) with the
FUV contours superposed (\cite{smi00}). The correspondence is excellent,
at least as far out
in the disk as the UIT data extend. We expect to see the FUV image and spiral
features grow further
when the GALEX data become available later this year, and we can
confidently predict that the close correspondence with the \HI\ will continue.

% ++++++++++++++++++++++++++++++++++++++++++++++++++++++++++++++++++++++
% ++++++++++++++++++++++++++++++++++++++++++++++++++++++++++++++++++++++

\subsection{\HI\ arcs and blisters}

The first study to successfully identify the characteristic PDR
``arc'' or ``blanket'' morphology of \HI\ in close association
with far-UV sources
in a nearby galaxy was carried out on M81 by \cite{all97}. The
problem is, of course, to obtain sufficient linear resolution ($\sim
100$ pc) in the \HI\ observations to permit one to identify the
morphology of the PDR structures. An important point to note is that
the best ``correlation'' is between the \HI\ and the far-UV,
and \textit{not} between the \HI\ and the H$\alpha$.
 
\begin{figure}[ht]
\centerline{\includegraphics[width=4in,height=3in]{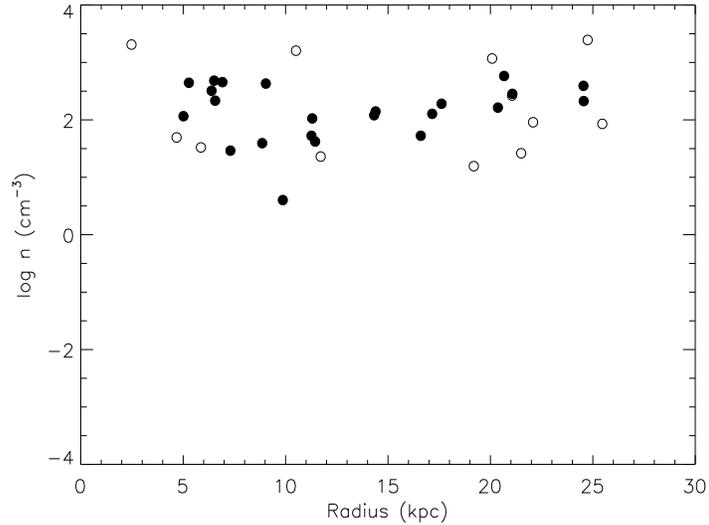}}
\caption{Total gas volume density in GMCs near a sample of
35 young star clusters in M101. This is all \Htwo\ deep within the cloud.
See Figure 19 in \cite{smi00} for further details. \label{fig:m101pdrs}}
\end{figure}

A study similar to that done on M81 but with more quantitative results has
been carried out on M101 by \cite{smi00}, who used VLA-\HI\ and
UIT far-UV data to identify and measure PDRs over the whole extent of
the M101 disk. From these observations they derived the volume density
of the \Htwo\ in the adjacent GMCs in the context of the PDR model.
Figure \ref{fig:m101pdrs} shows the best estimate of the \Htwo\ volume
densities of GMCs near a sample of 35 young star clusters.  The range
in density (30 - 1000 cm$^{-3}$) is typical for GMCs in our Galaxy,
lending support to the use of the PDR picture, and also shows little
trend with galactocentric distance.

It must be mentioned here that \cite{bra97} has offered a different
interpretation of the discrete \HI-bright features, which he called
the ``High-Brightness Network'' and identified with the ``Cold Neutral
Medium'' phase of the two-phase model for the ISM (\cite{fie69},
see also \cite{wol95} for a more recent discussion).
However, in a recent paper, \cite{wol03} favor the interpretation
of \cite{smi00} in terms of PDR-generated \HI\ .

% ++++++++++++++++++++++++++++++++++++++++++++++++++++++++++++++++++++++
% ++++++++++++++++++++++++++++++++++++++++++++++++++++++++++++++++++++++
There is other, IR spectral evidence that PDRs are important for
understanding the physics of the ISM in galaxy disks. KAO observations
of the $158\mu$m C\,{\sc ii} line suggest that as much as 70\%-80\% of
the \HI\ in NGC 6946 could be produced by photodissociation (\cite{mad93}),
and ISO spectra in the mid-IR indicate that the bulk of the
mid-IR emission from galaxy disks arises in PDRs (\cite{lau99};
\cite{vig99}).

\section{Some implications}

\begin{figure}[ht]
\sidebyside
{\includegraphics[width=1.8in, height=1.4in]{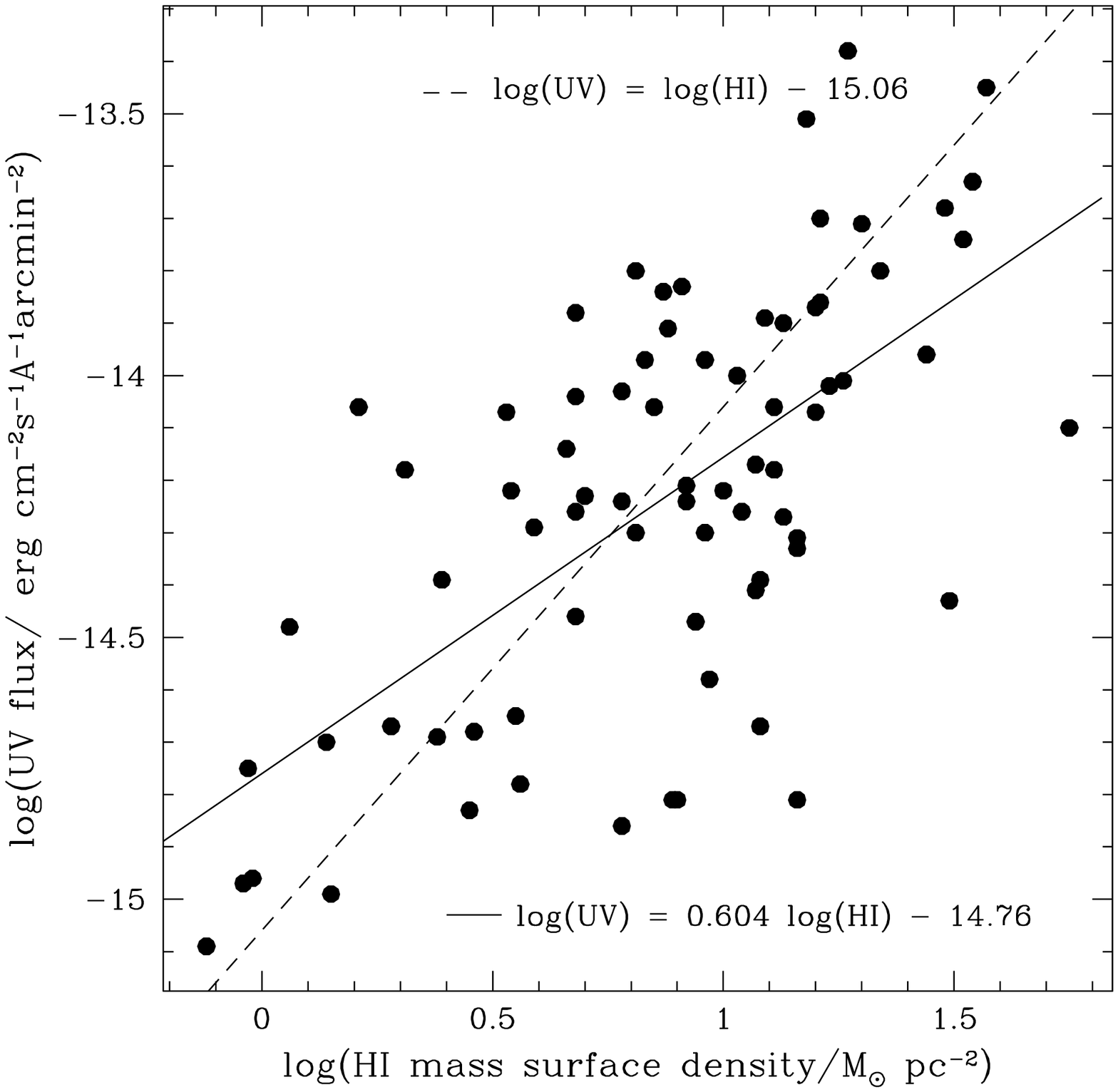}
\letteredcaption{a}{Data showing a correlation between
average observed 21-cm line surface brightness (converted to ``\HI\
mass surface density'') on the X-axis and the Far-UV surface brightness
(taken as a measure of the formation rate of massive stars) on the Y-axis,
from \cite{deh94}.  ``Schmidt Law'' fits are shown, with
indexes of 1 (dashed line) and 0.6 (solid line). \label{fig:SchmidtLawa}}}
{\includegraphics[width=1.8in, height=1.4in]{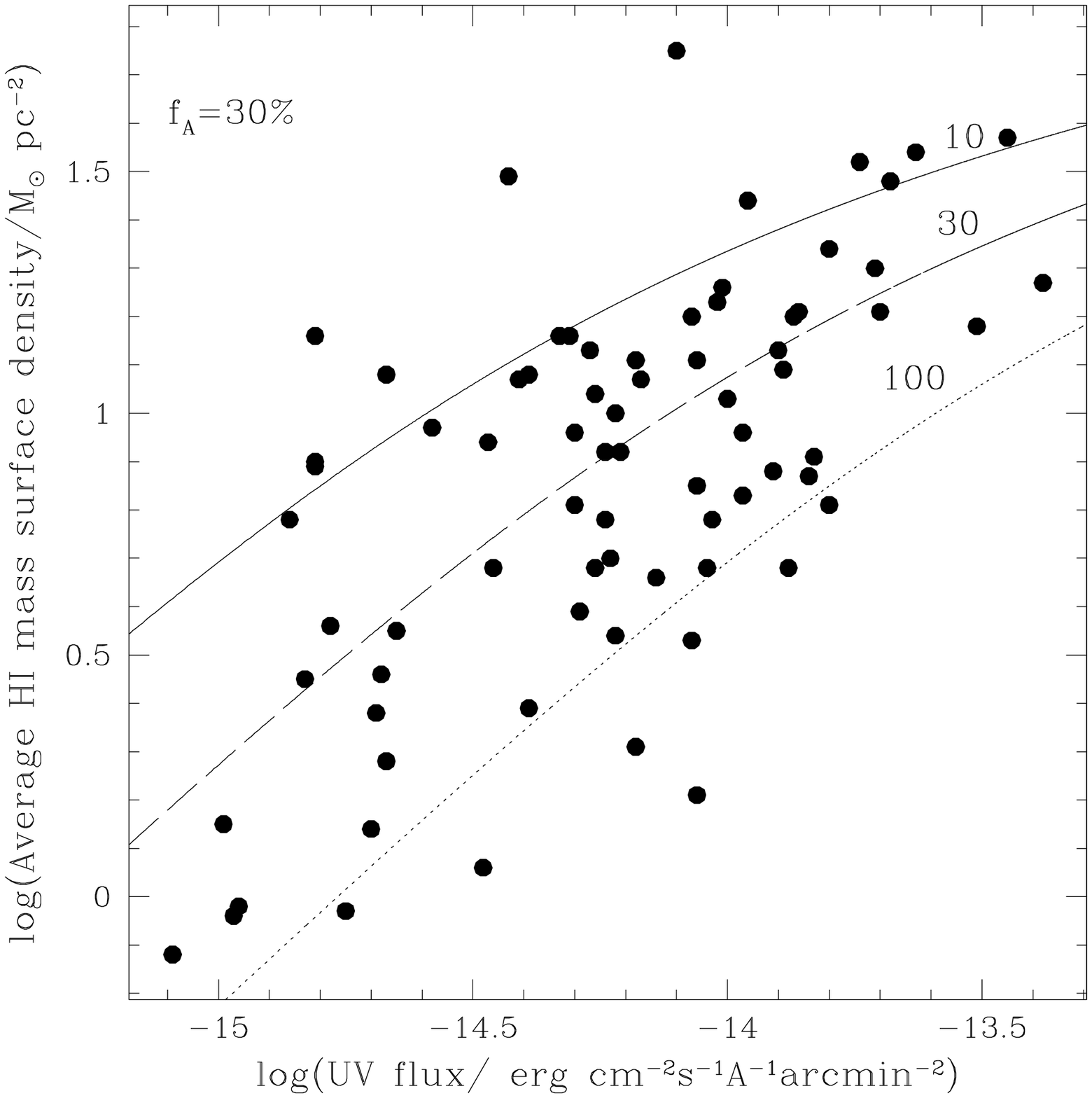}
\letteredcaption{b}{The data plotted with axes inverted so as to
emphasize the explanation in
terms of photodissociation.  The solid curves are the models of \HI\
production in PDRs described briefly in the text, and are labelled with the
proton volume densities of the parent GMCs.  The \HI\ area filling factor is
assumed to be 0.30 over the disk of the galaxy. \label{fig:SchmidtLawb}}}
\end{figure}

So, you say, let's suppose I am right in my view that \textit{\textbf{\HI\
is not the fuel for the star formation process in galaxies, but merely
the smoke from it!}} Why does this matter? Well, the idea that \HI\ directly and
quantitatively tracks the main component of the ISM in galaxy disks
is a basic tenet of our current view of star formation on the large scale in
galaxies, and it may take a few moments to consider the alternatives and the
consequences of ``shifting the paradigm''. I can think of at least 3
consequences at the moment:

\begin{itemize}
\item There must be significantly more gas present in galaxies in the form
of ``cold'' \Htwo. The \HI\ is showing us mainly only the surfaces of molecular
clouds, and the PDR model by itself does not provide a prescription for how to
go from the \HI\ as a surface phenomena to the \Htwo\ in the
volume of the clouds. But we can confidently predict that \textit{more} \Htwo\
will be present than we currently think. \Htwo\ in amounts from 2 - 5 times that
of the known \HI\ could probably be ``hiding'' in galaxy disks without clearly
violating any known constraints.
\item The far outer parts of galaxy disks, where small amounts of \HI\ appear
with only sparse star formation, are prime sites for ``hiding'' \Htwo. A focussed
effort to find such gas ought to be made. Possibilities for detection may
include measurements of dust opacity (it is likely to be too cold to emit any
appreciable amounts of Far-IR continuum or line emission, but this needs
to be considered carefully\footnote{The paper by Boulanger at this meeting
has made a start in this regard.}), and molecular absorption lines. The
anomalous absorption of the Cosmic Microwave Background by the 6 and 2-cm lines
of formaldehyde \Htwo CO are intriguing possibilities which ought to be explored
further.
\item We are interpreting the ``Schmidt Law'' for global star formation
\textit{backwards}. This is a particularly far-reaching consequence of the
photodissociation picture favored here. The situation has been described
by \cite{all02}.
Basically, we have inverted ``cause''and ``effect'' for many years in this discussion, viewing the \HI\ column as the cause, and the star formation rate
(e.g.\ quantified by the FUV flux) as the effect. The observed relationship
between these two quantities (roughly a power law on a log-log plot) is called
the ``Schmidt Law for Global Star Formation'', and has been the basis for many
attempts to develop a physical theory for large-scale star formation in galaxies
involving gravitational instablility in the disk. Such a
theory is still incomplete. On the other hand, the PDR picture
favored here views the \textit{FUV flux as the cause} and the
\textit{\HI\ column as the effect}, and provides a simple
explanation for the observed correlation in terms of physics we
already know (see Figures \ref{fig:SchmidtLawa} and \ref{fig:SchmidtLawb},
from \cite{all02}).
\end{itemize}

% ++++++++++++++++++++++++++++++++++++++++++++++++++++++++++++++++++++++
% ++++++++++++++++++++++++++++++++++++++++++++++++++++++++++++++++++++++

\section{Summary Remarks}

To summarize the latest views on this topic, first a conclusion
which has been corroborated by several authors and which by now
seems quite solid:

\begin{itemize}
\item The \HI\ spiral arms in the inner parts of ``grand design'' galaxies
consist mostly (and perhaps even entirely) of photodissociated \Htwo.
\end{itemize}

To this I would add the following points established in papers by myself
and my co-workers:

\begin{itemize}
\item As well as O stars, B stars born in spiral arms play a major
role in this process;
\item The PDR morphological signature is widespread in HI when enough
linear resolution ($\lesssim 100$ pc) is available, and;
\item since there seems to be no reason to have more than one \HI\ formation
mechanism, I conclude that both the inner and the (far) outer \HI\ arms in
spirals are photodissociated \Htwo.
\end{itemize}

New results on the radial distributions of \HI\ described in this review are tantalizing, but need further work:

\begin{itemize}
\item The general shape of the \HI\ radial distributions in many spirals appears
to be amenable to explanation in the context of a simple photodissociation model:
	\begin{itemize}
	\item variations in \Htwo\ density, FUV intensity, and dust/gas ratio
	can control the appearance of \HI, and;
	\item Additional \HI\ may be produced by an elevated flux of cosmic rays
	in some galaxies.
	\end{itemize}
\end{itemize}

Photodissociation of \Htwo\ can explain a number of features in the morphology
of \HI\ in galaxies on scales from 100 pc to 10 kpc. Note that this process
is bound to produce \HI\ as long as
\Htwo\ and FUV photons are present; the real question we need to answer is
``how much?'', i.e. what
fraction of the total \HI\ content in a galaxy is cycling repeatedly through
a molecular phase on time scales which are short compared to the
dynamical evolution time of the
galaxy? If this fraction proves to be large, then there must be an even larger
reservoir of \Htwo\ present to sustain it. Estimating the quantities of
cool/cold \Htwo\ hiding in
the ISM of disk galaxies is pure guesswork at
the moment, but factors of 2 - 5 times
that in the form of \HI\ may not be unreasonable, and with a spatial distribution
as extended as the \HI\ itself.

\begin{acknowledgments}
I am grateful to my colleagues at STScI for their contributions to the
stimulating scientific environment we enjoy there, to Hal Heaton and
Michael Kaufman for their collaboration on the models described here,
and to David Block and others on the SOC for the opportunity to attend
this meeting and to present my views on the subject of \HI\ in galaxies.
\end{acknowledgments}

\chapappendix{Derivation of the \HI\ column density in equation \ref{eqn:nhi}}

Equation \ref{eqn:nhi} is obtained by setting the rate of formation of \Htwo\ on grains equal to the rate of destruction through photodissociation by far-UV
photons. Using the notation of \S \ref{sec:HI}:
\begin{eqnarray}
R_{form} \times n \times n_{1} & = & R_{diss} \times n_{2} \\
\nonumber  & = & D G_0 \times e^{-\tau_{gr,1000}} \times f_s(N_2) \times n_2 .
\end{eqnarray}
For a simple 1D slab geometry at constant density $n = n_1 + 2n_2$,
we can write $n_1 = dN_1/dx$, etc., so that the equilibrium equation becomes:
\begin{equation}
\nonumber R_{form} \times n \times (dN_1/dx)  =  D G_0 \times
e^{-\sigma (N_1 + 2N_2)} \times f_s(N_2) \times (dN_2/dx).
\end{equation}
This can be written as a simple first-order separable differential equation in the
column densities:
\begin{equation}
\nonumber e^{+\sigma N_1} \times dN_1  = 
\frac{D G_0}{R n}f_s(N_2)e^{-2 \sigma N_2} \times dN_2.
\end{equation}
This can be integrated through the entire slab on the half-plane $x \geq 0$
to give:
\begin{equation}
e^{+\sigma N_1}\vert_{x = \infty} - e^{+\sigma N_1}\vert_{x = 0} =
\frac{D G_0}{R n} \int_0^{\infty} f_s(N_2)e^{-2 \sigma N_2} \sigma dN_2. 
\end{equation}
The integral over $N_2$ on the RHS is just some number, call it $\mathcal{G}$,
so that:
\begin{equation}
N_1 = \frac{1}{\sigma} \ln [1 + \frac{D G_0 \mathcal{G}}{R n}].
\end{equation}

% End main text and appendices

\end{document}